\documentclass[aps,twocolumn,showpacs,amsmath,amssymb,amsfonts]{revtex4}
\usepackage{graphicx}
\usepackage{bm}
\newcommand{\tst}{\textstyle}

\newcommand{\mbf}{\mathbf}
\newcommand{\mrm}{\mathrm}
\newcommand{\ud}{\mathrm{d}}

\begin{document}

\author{Z. Idziaszek$^{1,2}$ and T. Calarco$^{1,3}$}
\affiliation{$^1$Istituto Nazionale per la Fisica della Materia,
BEC-INFM Trento, I-38050 Povo (TN), Italy\\
$^2$Centrum Fizyki Teoretycznej, Polska Akademia Nauk, 02-668
Warsaw, Poland\\
$^3$European Centre for Theoretical Studies in Nuclear Physics and
Related Areas, I-38050 Villazzano (TN), Italy}

\title{Two atoms in an anisotropic harmonic trap}

\begin{abstract}
We consider the system of two interacting atoms confined in axially symmetric harmonic trap. Within the pseudopotential
approximation, we solve the Schr\"{o}dinger equation exactly, discussing the limits of quasi-one and quasi-two-dimensional geometries. 
Finally, we discuss the application of an energy-dependent pseudopotential,
which allows to extend the validity of our results to the case of tight traps and large scattering lengths.
\end{abstract}

\pacs{34.50.-s, 32.80.Pj}

\maketitle

Atomic interactions at ultralow temperatures are of central
importance for recent research on quantum degenerate gases \cite{Dalfovo}.
A typical feature of experiments on ultracold matter is the presence 
of a weak trapping potential, which modifies the properties of the 
cloud of atoms, while it does not affect the collisions of individual particles. 
Development of optical lattice technology, however, has created
systems where the atoms are tightly confined in the wells of 
optical potential \cite{Bloch}.
In addition, the experimental achievement of the Mott insulator phase
\cite{Greiner} has allowed for a precise control over a number of
atoms stored in a single well. This has opened a way for
experimental studies of interactions of individual atoms in the
presence of trapping potential and, together with other approaches
to micromanipulation of neutral atoms like atom chips
\cite{Schmiedmayer,Birkl} or tight dipole traps \cite{Grangier},
it represents a major candidate for the implementation of quantum
information processing. A theoretical understanding of the dynamics of
few atoms in deformed tight-confining geometries would be of great 
help in all these contexts.

From the theoretical side, the analytical solution for two atoms
interacting in a harmonic trap is known only for the spherically
symmetric case \cite{Busch,Block}. The corresponding problem for
axially symmetric trap was studied numerically in \cite{BoldaQ}.
However, there the authors considered only the limiting regimes of
quasi-one and quasi-two-dimensional traps. In this letter we
present the exact solution for the axially symmetric harmonic trap
of arbitrary geometry. In particular, when the ratio of axial to
radial trapping frequency is an integer, or the inverse of an
integer, we give the explicit analytic form of the exact solution.
In the other cases, we derive an efficient recurrence relation
that allows for evaluating it. Furthermore, we study the asymptotic
behavior of eigenenergies and eigenfunctions in the limit of quasi-one
and quasi-two-dimensional traps.

A standard treatment of ultracold atom interactions is based on
the replacement of a real physical potential by an $s$-wave
delta-function pseudopotential.  To
extend the validity of this model interaction to the case of 
tight traps and large scattering lengths, one can utilize
the concept of an effective, energy-dependent scattering length
\cite{Bolda}. We discuss this idea and show how our results 
can be generalized to the case of magnetically tunable Feshbach 
resonances.

We consider two interacting atoms of mass $m$ confined in an axially symmetric
harmonic trap with frequencies $\omega_{\perp}$ and $\omega_{z}$. In the
following we use dimensionless variables, in which all lengths are
expressed in units of harmonic oscillator length
$a_z=\sqrt{\hbar/m \omega_z}$, and all energies are expressed in
units of $\hbar \omega_z$. In these units the trapping potential is
$V_{T}(\mathbf{r}) = \frac{1}{2} ( \eta^2 \rho^2 + z^2 )$,
where $\eta = \omega_{\perp}/\omega_z$ and $\rho^2 = x^2 + y^2$.
We assume that the range of the
interatomic potential is much smaller than the oscillator lengths
$a_z$ and $a_{\perp}=\sqrt{\hbar/m \omega_{\perp}}$, which guarentees 
that the
interatomic potential is not distorted by the harmonic trap. For
sufficiently low energies, the scattering is purely of $s$-wave
type and we model the atom-atom interaction by a Fermi
pseudopotential $V(\mathbf{r})=4 \pi a \delta(\mathbf{r})
\frac{\partial}{\partial r} r$ with $s$-wave scattering length
$a$ \cite{Fermi}. For the harmonic confining potential, the total Hamiltonian 
\begin{equation}
\label{Htot}
\hat{H} =  - \frac{1}{2} \nabla^2_1 - \frac{1}{2} \nabla^2_2 
+ V_{T}(\mathbf{r}_1) + V_{T}(\mathbf{r}_2) + V(\mathbf{r}_1 - \mathbf{r}_2),
\end{equation}
can be splitted 
into center of mass part: $\hat{H}_\mrm{CM} = - \frac{1}{2} \nabla^2_{R} + V_{T}(\mathbf{R})$, and the relative motion part: $\hat{H}_\mrm{rel} =  - \frac{1}{2} \nabla^2_{r} + V_{T}(\mathbf{r}) + V(\sqrt 2 \mathbf{r})$,
where $\mbf{r}= (\mbf{r}_1 - \mbf{r}_2)/\sqrt{2}$ and 
$\mbf{R}= (\mbf{r}_1 + \mbf{r}_2)/\sqrt{2}$. To solve the 
Schr\"odinger equation for the relative motion, 
we decompose the wave function in
the basis of eigenstates of the noninteracting problem, substitute
this decomposition into the Schr\"odinger equation, and then
extract the expansion coefficients by projecting onto
noninteracting states \cite{Busch}. This yields the wave function
of $m_z = 0$ states, with vanishing angular momentum along
$z$-axis
\begin{align}
\label{Psi} \Psi(\mbf{r}) = & \frac{\eta}{(2 \pi)^{\frac{3}{2}}}
\int_{0}^{\infty} \! \! \! \! \ud t \ \frac{\exp\!\left[ t E
-\frac{z^2}{2} \coth t - \frac{\eta \rho^2}{2} \coth(\eta t)
\right]} {\sqrt{\sinh(t)} \sinh (\eta t)}.
\end{align}
The harmonic oscillator states with $m_z \neq 0$ vanish at $\mbf{r} = 0$,
and they are not influenced by the pseudopotential.
Eq.~(\ref{Psi}) represents the wave function which is not normalized, and is related with the
single particle Green function of the anisotropic harmonic oscillator by $\Psi(\mbf{r}) = - 2 G(\mbf{r},0)$.
We note that the integral representation (\ref{Psi}) is valid for energies smaller than
the ground state energy of the harmonic oscillator: $E_0=1/2+\eta$. The validity of Eq.~(\ref{Psi}), however,
can be extended for $E \geq E_0$ by means of the analytic continuation.

The presence of the trapping potential implies the discrete
character of the energy spectrum. The allowed values of energy $E$
has to be determined from equation: $- 1/\left(\sqrt{2} \pi
a\right) = \left[ (\partial/\partial r) r \Psi(\mbf{r})
\right]_{r=0}$, which results from derivation of Eq.~(\ref{Psi}),
and expresses a boundary condition imposed by
zero-range interaction. Investigation of the integral in
Eq.~(\ref{Psi}) for small values of $\mbf{r}$, shows that
$\Psi(\mbf{r})$ behaves like $1/(2 \pi r)$ as $\mbf{r} \rightarrow
0$. This divergence is removed by the regularization
operator $(\partial/\partial r) r$ in the Fermi pseudopotential.
Subtracting from the integral (\ref{Psi}), the part which gives
rise to the $1/r$ singularity, the condition for the eigenenergies can
be rewritten as
\begin{equation}
\label{Energ}
- \sqrt{2 \pi}/a = {\cal F}\left(- (E-E_0)/2,\eta \right),
\end{equation}
where
\begin{equation}
\label{Deff}
{\cal F}(x,\eta) \equiv
\int_{0}^{\infty} \! \! \! \! \ud t \ \left[ \frac{ \eta e^{- x t}}
{\sqrt{1-e^{-t}} \left(1 - e^{-\eta t}\right)} - \frac{1}{t^{3/2}} \right].
\end{equation}
For particular values of the anisotropy parameter $\eta$, the function ${\cal F}(x,\eta)$ can be calculated
analytically. In the case of cigar shaped traps with $\eta=n$, where $n$ is a positive integer, we obtain
\begin{equation}
{\cal F}(x,n) =  \label{FCig}\frac{\sqrt{\pi}\,\Gamma(x)\!
\displaystyle\sum_{m=1}^{n-1} F\!\left(1,x;x+{\tst
\frac{1}{2}};e^{i \frac{2\pi m}{n}}\right)}{\Gamma(x+\tst\frac
12)}-\frac{2\sqrt{\pi}\, \Gamma(x)}{\Gamma(x-\textstyle\frac 12)},
\end{equation}
where $F(a,b;c;x)$ denotes the hypergeometric function and $\Gamma(x)$ is the Euler gamma function.
It can be easily verified that the sum in Eq.~(\ref{FCig})
involving complex roots of unity is a real number for $x \in \mathbf{R}$. On the other hand,
for pancake shaped traps with anisotropy parameter $\eta=1/n$, the following result holds
\begin{equation}
\label{FPan} {\cal F}(x,1/n) = - \frac{2 \sqrt{\pi}}{n}
\sum_{m=0}^{n-1} \frac{\Gamma(x+m/n)}{\Gamma(x-1/2+m/n)}.
\end{equation}
For $n=1$, we recover obviously the well known result for the spherically symmetric trap:
${\cal F}(x,1) = - 2 \sqrt{\pi} \; \Gamma(x)/\Gamma(x-1/2)$ \cite{Busch}.
We note that Eqs.~(\ref{FCig}) and (\ref{FPan}) are derived from the integral representation (\ref{Deff})
applicable for $x>0$, however, their validity for $x<0$ is extended by virtue of the analytic continuation.

In the general case, when $\eta$ does not meet the conditions of Eqs.~(\ref{FCig}) and (\ref{FPan}),
the energy spectrum can be determined numerically. For $E<E_0$ the function ${\cal F}(x,\eta)$ is given by
by Eq.~(\ref{Deff}), while for $E>E_0$, one can utilize the following recurrence relation
\begin{equation}
\label{FDiff} {\cal F}(x,\eta)-{\cal F}(x+\eta,\eta) = \eta
\sqrt{\pi} \, \Gamma(x)/\Gamma(x+1/2),
\end{equation}
which can be easily derived from the definition of ${\cal F}(x,\eta)$.

From the practical point of view, the use of the exact results of
Eqs.~(\ref{FCig}) and (\ref{FPan}) is efficient as long as $n$ is
not too large. To determine the energy levels in the limit of
quasi-one- and quasi-two-dimensional traps, we derive the
asymptotic form of ${\cal F}(x,\eta)$ for $\eta \gg 1$ and $\eta
\ll 1$.

Let us first focus on the case of $\eta \gg 1$. Performing an expansion
in the integral (\ref{Deff}) for large $\eta$ and making use of 
the recurrence formula (\ref{FDiff}) we arrive at
\begin{equation}
\label{FQ1D} {\cal F}(x,\eta) \stackrel{\eta \gg 1}{\approx} \sqrt{\pi \eta}\,[\zeta
(\textstyle\frac 12,1+x/\eta) + \sqrt{\eta}\,
\Gamma(x)/\Gamma(x+\textstyle\frac 12)],
\end{equation}
where $\zeta(s,a)$ denotes the Hurwitz zeta function. This asymptotic
formula is valid for $x > - \eta $, which corresponds to the range of
energies $E< E_0+ 2 \eta$. For the lowest excited states: $0<E-E_0
\ll 2 \eta$ we approximate $\zeta(1/2,1+x/\eta)$ by
$\zeta(1/2,1)$ in Eq.~(\ref{FQ1D}),
and match the resulting energy spectrum with the
energy spectrum of two atoms in a one-dimensional trap. The latter
is determined by $\sqrt{2} a_{1D} =
\Gamma\left((E_0-E)/2\right)/\Gamma\left((E_0+1-E)/2 \right)$
\cite{Busch}. The two spectra are identical, provided that the
one-dimensional scattering length is $a_\mrm{1D} = -1 /\eta a -
\zeta(1/2,1)/\sqrt{2 \eta}$, which agrees with the value of
the re\-nor\-ma\-li\-zed scattering length derived for a
quasi-one-dimensional waveguide \cite{Olshanii}. On other hand, for
energies $E<E_0$, we can use ${\cal F}(x,\eta) \approx \sqrt{\pi \eta} \, \zeta (1/2,x/\eta)$, which
follows from Eqs.~(\ref{FDiff}) and (\ref{FQ1D}). This
approximation, substituted into (\ref{Energ}), leads to the
condition determining the energy of a bound state:
\begin{equation}
\label{Ebs1D}
\sqrt{2}/a  + \sqrt{\eta} \zeta \left(1/2,(E_0-E)/(2\eta)\right)=0,
\end{equation}
which is identical to the known result derived for the
quasi-one-dimensional waveguide \cite{Olshanii,Bergeman}.

In the case of quasi-two-dimensional traps: $\eta \ll 1$, we obtain the following approximate
formula for ${\cal F}(x,\eta)$:
\begin{equation}
\label{FQ2D}
{\cal F}(x,\eta) \stackrel{\eta \ll 1}{\approx} - \Phi (x) - \log (\eta) - \psi \left(x/\eta\right),
\end{equation}
where
\begin{align}
\label{PhiSer}
\Phi (x) = & \: \: 2 - \log (1+x)  \\
\nonumber& {} + 2 \sum_{k=1}^{\infty} \frac{(2k)!}{(2^k k!)^2}
\left[(k+{\tst \frac{1}{2}})\log\!\frac{x+k}{x+k+1}+1\right],
\end{align}
and $\psi(z) = (d/dz) \log\Gamma (z)$ denotes the digamma function. 
This result is valid for $x>-1$, which corresponds to
energies $E<E_0+2$. For the lowest excited states: $0<E-E_0 \ll
2$, we approximate $\Phi (x)$ by  $\Phi(0)$ in Eq.~(\ref{FQ2D}),
and compare the resulting energy spectrum to that of the
two-dimensional system. In the two-dimensional trap, the
eigenenergies of two interacting atoms are given by $- \log ( 2
a_\mrm{2D}^2 \eta) = \psi\left((E_0-E)/(2\eta)\right)$
\footnote{This result assumes that
the two-dimensional scattering length $a_\mrm{2D}$ is related to 
the {\it s}-wave phase shift $\delta_0$
by $\tan \delta_0 = (\pi/2) \log^{-1} (k a_{2D})$, with $k^2 = E$.}.
In this way we find the value of the two-dimensional
scattering length $a_\mrm{2D}$ for which both spectra are the
same: $a_\mrm{2D} =\exp[\frac 12({\cal D} -
\sqrt{2\pi}/a_\mrm{3D})]/\sqrt 2$, where ${\cal D} = \Phi(0)
\simeq 1.938$. This result agrees with the value of $a_\mrm{2D}$
derived for a quasi-two-dimensional system without confinement in
the radial direction \cite{Petrov}. In the range of energies
corresponding to a bound state, we use an asymptotic expansion of
$\psi(x/\eta)$ for $ x / \eta \gg 1$ in Eq.~(\ref{FQ2D}), which
yields ${\cal F}(x,\eta) \approx - \Phi (x) - \log x$.
Substituting this approximation into (\ref{Energ}), we obtain the
equation which determines the energy of a bound-state in
quasi-two-dimensional traps:
\begin{equation}
\label{Ebs2D}
\sqrt{2 \pi}/a = \Phi((E_0-E)/2) + \log((E_0-E)/2).
\end{equation}
For a shallow bound-state ($E_0 -E \ll 1$) one can approximate 
$\Phi((E_0-E)/2)$ by $\Phi(0)$ and in this regime the binding energy is
given by $E_0-E= 0.288 \exp(\sqrt{2 \pi}/a)$ \cite{Petrov}.

\begin{figure}
     \includegraphics[width=86mm]{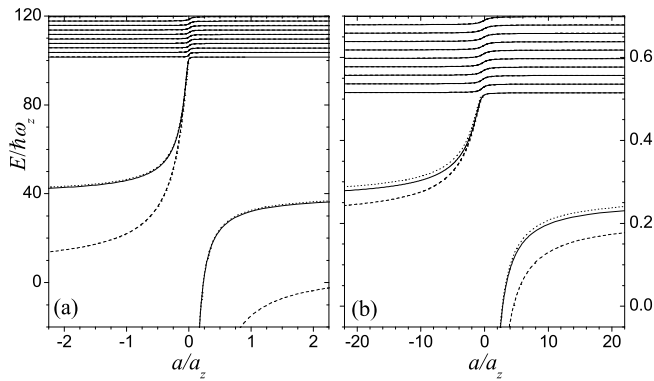}
     \caption{
     \label{fig:En}
     Energy spectrum of two atoms interacting via regularized delta potential in a
     three-dimensional trap with $\eta = \omega_{\perp}/\omega_z = 100$ (a) and
     $\eta = 0.01$ (b). Panel (a): The exact energy levels (solid lines) are
     compared with the energy spectrum of the one-dimensional system with renormalized scattering length (dashed lines),
     and with the energies of a bound state calculated from Eq.~(\ref{Ebs1D}).
     (dotted lines - almost indistinguishable from the solid ones).
     Panel (b): The exact energy levels (solid lines) are
     compared with the energy spectrum of the two-dimensional system with renormalized scattering length (dashed lines),
     and with the energies of a bound state calculated from Eq.~(\ref{Ebs2D})
     (dotted lines). The scattering length $a$ is scaled in h.o. units $a_z = \sqrt{\hbar / m \omega_z}$.
     }
\end{figure}

Fig.~\ref{fig:En} shows the energy spectrum of two interacting atoms calculated for $\eta=100$ (a)
and $\eta=0.01$ (b). Fig.~\ref{fig:En}(a)
compares the exact energy levels given by Eqs. (\ref{Energ}) and (\ref{FCig}),
with the energy spectrum of the one-dimensional system with renormalized scattering length $a_\mrm{1D}$,
and with bound-state energies calculated from Eq.~(\ref{Ebs1D}). Fig.~\ref{fig:En}(b)
presents the exact result of Eqs. (\ref{Energ}) and (\ref{FPan}),
the energy spectrum of the two-dimensional system with renormalized scattering length $a_\mrm{2D}$,
and bound-state energy calculated from Eq.~(\ref{Ebs2D}).
We have not included the energy levels calculated from approximations (\ref{FQ1D}) and
(\ref{FQ2D}), which for $\eta=100$ and $\eta=0.01$ are indistinguishable from the exact result.
We observe that for $E>E_0$ the one- and the two-dimensional spectra fit very well the exact eigenenergies,
whereas they are incorrect with respect to the bound-state energies. The latter, however, are well
described by Eqs.~(\ref{Ebs1D}) and (\ref{Ebs2D}).
%%%%%%%%%%%%%%%%%% Figure 1 %%%%%%%%%%%%%%%%%%%%%%
%%%%%%%%%%%%%%%%%%%%%%%%%%%%%%%%%%%%%%%%%%%%%%%%%%%

We now turn to the calculation of wave functions. While for $E<E_0$ they can be evaluated from 
the integral representation
(\ref{Psi}), in the general case, they can be determined from the following expansions
\begin{align}
\nonumber
\Psi(\mbf{r}) =  \frac{\eta \, e^{-\eta \rho^2/2}}{2 \pi^{3/2} 2^{{\cal E}/2}}
\sum_{m=0}^{\infty} & \left[ 2 ^{\eta m} \Gamma({\tst \frac{2 \eta m -{\cal E}}{2}})
L_m(\eta \rho^2) \right. \\
\label{PsiExp1}
& \, \times \left. D_{{\cal E}-2 \eta m}(|z|\sqrt{2}) \right], \\
\nonumber
\Psi(\mbf{r}) = \frac{e^{-(\eta \rho^2+ z^2)/2}}{2 \pi^{3/2}} \sum_{k=0}^{\infty} &
\left[ \frac{(-1)^k}{2^{2k} k!} H_{2k}(z)
\Gamma\!\left({\tst \frac{k}{\eta} -\frac{{\cal E}}{2 \eta}}\right) \right. \\
\label{PsiExp2}
& \, \times \left.
U\!\left({\tst \frac{k}{\eta} -\frac{{\cal E}}{2 \eta}},1,\eta \rho^2\right)
\right].
\end{align}
Here ${\cal E}=E-E_0$, $L_m(x)$ and $H_k(z)$ are respectively the Laguerre and Hermite polynomials, 
$D_{\nu}(x)$ is the parabolic cylinder function and $U(a,b,z)$ 
denotes the confluent hypergeometric function. 
As it can be easily observed, the first expansion involves the harmonic oscillator 
wave functions in the radial direction and the one-dimensional
solution for two interacting atoms in the axial direction. We have verified that for elongated traps ($\eta \gg 1$),
the first term of this series
provides already a quite good approximation for the wave function of the lowest excited states.
A similar feature is observed for
the second series in Eq.~(\ref{PsiExp2}) in the traps with $\eta \ll 1$. Conversely, for energies
$E<E_0$ the two series involve generally several terms. In this regime, we can analyze the behavior
of the wave functions on the basis of the integral representation (\ref{Psi}). Due to the complicated form
of the latter integral, we focus here only on the limiting case of quasi-one- or quasi-two-dimensional traps, and
investigate only the behavior of the axial ($\rho =0$) and the radial ($z=0$) profiles of the wave functions.

Expanding the integral in Eq.~(\ref{Psi}) for $\eta \gg 1$, we obtain the axial
($\Psi_z(z) \equiv \Psi(z \hat{\mbf{z}})$)
and radial ($\Psi_\perp(\rho) \equiv \Psi(\rho \hat{\bm{\rho}})$) profiles of the wave function,
applicable for $E<E_0$
\begin{align}
\Psi_z(z) \stackrel{\eta \gg 1}{\approx} & \frac{\eta}{2 \pi}
\sum_{m=0}^{\infty} \frac{\exp \left(-2 |z|\sqrt{m \eta-{\cal
E}/2}\,\right)}{\sqrt{m \eta-{\cal E}/2}},
\label{PsiQ1D_z} \\
\Psi_{\perp}(\rho) \stackrel{\eta \gg 1}{\approx} & e^{-\eta \rho^2/2}
\left[\rho^{-1} + \sqrt{\eta} \, \zeta \! \left({ \tst
\frac{1}{2},-\frac{{\cal E}}{2 \eta}} \right)\right]/(2\pi).
\label{PsiQ1D_r}
\end{align}
For $|z|\sqrt{-{\cal E}}\gg 1$, the main contribution to the sum in Eq.~(\ref{PsiQ1D_z})
comes from the first term. In this case the wave function exhibits the exponential decay,
which is similar to the behavior of one-dimensional bound-state in a free space: $\Psi(z) \sim \exp(-\sqrt{-2 {\cal E}} |z|)$. 
On the other hand, the wave function in the radial direction has a Gaussian profile, characteristic for the
ground-state of harmonic oscillator, whereas the divergent term $1/(2\pi \rho)$ arises due to the interaction
potential.

In quasi-two-dimensional traps, for energies $E<E_0$, we found the following radial and axial profiles of the wave-functions
\begin{align}
\Psi_{\perp}(\rho) \stackrel{\eta \ll 1}{\approx} & \pi^{-\frac{3}{2}}
\sum_{m=0}^{\infty} \frac{(2m)!}{(2^m m!)^2} K_0 ({\tst 2 \rho
\sqrt{m -{\cal E}/2 }\,}),
\label{PsiQ2D_r} \\
\Psi_{z}(z) \stackrel{\eta \ll 1}{\approx} & \frac{e^{-
z^2/2}}{2\pi} \left[ \frac{1}{|z|} - \frac{\Phi(-{\cal E}/2) +
\log (- {\cal E}/2)} {\sqrt\pi} \right], \label{PsiQ2D_z}
\end{align}
where $K_0(x)$ is a modified Bessel function. The asymptotic behavior of $K_0(x)$ for $x\gg1$ is governed
by $K_0(x)\sim \sqrt{\pi/2x} e^{-x}$. Hence, for $\rho\sqrt{-{\cal E}}\gg 1$, the sum in
(\ref{PsiQ2D_r}) is dominated by the first term, and the asymptotic decay of the wave function in the 
radial direction is
similar to the one observed for a bound state in two dimensions:
$\Psi(\rho) \sim K_0( \sqrt{-2 {\cal E}}\rho)$. Along the tightly confined, axial direction, the wave
function has a Gaussian profile, which is modified at short distances by the interaction potential.

\begin{figure}
   \includegraphics[width=86mm]{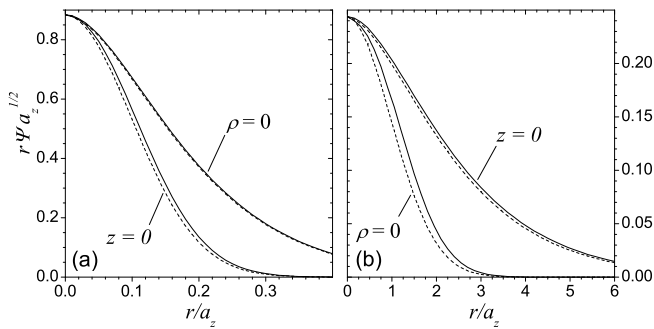}
     \caption{
     \label{fig:Cut}
     The axial ($\rho=0$) and the radial ($z=0$) profiles of the ground-state wave function for two atoms interacting
     via a regularized delta potential with $a=\pm \infty$. The atoms are confined in a harmonic trap
     with $\eta = \omega_{\perp}/\omega_z = 100$ (a), and $\eta = 0.01$ (b).
     The exact profiles (solid lines) are compared with the approximate results of Eqs.
     (\ref{PsiQ1D_z})-(\ref{PsiQ2D_r}) (dashed lines). All lengths are scaled to $a_z = \sqrt{\hbar / m \omega_z}$.
     }
\end{figure}

The behavior of the ground-state wave function in the unitarity limit ($a=\pm \infty$) in 
the quasi-one-dimensional ($\eta=100$) and quasi-two-dimensional traps ($\eta=0.01$)
is presented in Fig.~\ref{fig:Cut}. The figure compares the exact profiles evaluated from 
Eqs.~(\ref{PsiExp1}) and (\ref{PsiExp2}) with the approximate results of Eqs.~(\ref{PsiQ1D_z})-(\ref{PsiQ2D_r}).
We observe that all approximate curves fit quite well the exact functions.
%%%%%%%%%%%%%%%%%% Figure 2 %%%%%%%%%%%%%%%%%%%%%%%%%%%%%%%%%%
%%%%%%%%%%%%%%%%%%%%%%%%%%%%%%%%%%%%%%%%%%%%%%%%%%%

Finally we would like to stress that our derivation can be easily
supplemented to include an energy-dependent scattering length
\cite{Bolda,Blume,Stock}. 
This extends the validity of the
pseudopotential approximation to scattering lengths much larger than
the trap size, and allows to properly describe the entire
molecular spectrum. The energy-dependent effective scattering
length is defined through the $s$-wave phase shift $\delta_0$:
$a_{\mrm{eff}}(E) = - \tan \delta_0 (k)/k$, where $\hbar k$ is the
relative momentum \footnote{ For negative energies $a_{\mrm{eff}}(E)$ 
can be defined by analytic continuation. For details see \cite{Stock}.}.
The application of this model
interaction in our derivations leads to substitution of $a$ by
$a_{\mathrm{eff}}(E)$ in Eq.~(\ref{Energ}) determining the eigenenergies,
and requires a self-consistent solving for the value of $E$. For 
magnetically tunable Feshbach resonances, 
the $s$-wave phase shift is known analytically \cite{TheoryFeshbach}, 
and in this case one can derive an explicit formula for 
$a_{\mathrm{eff}}(E)$ \cite{Bolda}.

In summary, we solved analytically the problem of two atoms interacting in an
axially symmetric harmonic trap with arbitrary trap anisotropy.
For integer ratios of the trapping frequencies we gave closed
formulas for the solutions. Furthermore, by introducing an
effective energy dependence in the scattering length \cite{Bolda,Blume},
we can find the solutions for any value of the latter. Therefore
our result allows for a direct exact evaluation of the dynamics of
a pair of interacting neutral atoms in very tight traps, possibly
in reduced dimensionality and under an arbitrary external magnetic
field, even in the presence of Feshbach resonances. Applications
include a significant range of situations involving quantum
control at the atomic level, from single-atom interferometry to
quantum information processing.

We thank L.P. Pitaevskii, M. Holland, G. Orso and M. Wouters for valuable discussions.
We are grateful to E. Bolda for making available his numerical data reported 
in \cite{BoldaQ}. T. Calarco acknowledges support from the EC under contract
IST-2001-38863 (ACQP) and from MIUR (COFIN 2002).

\end{document}